\documentclass[twocolumn,preprintnumbers,amsmath,amssymb,superscriptaddress]{revtex4}

\usepackage{graphicx}
\usepackage{times}
\usepackage{amssymb}

\newcommand{\be}{\begin{equation}}
\newcommand{\ee}{\end{equation}}
\newcommand{\bea}{\begin{eqnarray}}
\newcommand{\eea}{\end{eqnarray}}

\newcommand{\Ket}[1]{\vert \, #1 \, \rangle}
\newcommand{\Bra}[1]{\langle \, #1 \,\vert}

\renewcommand{\epsilon}{\varepsilon}

\begin{document}


\title{Amplitude Spectroscopy of a Solid-State Artificial Atom}

\author{D.~M.~Berns}
 \affiliation{Department of Physics, Massachusetts Institute of Technology, Cambridge MA 02139}
\author{M.~S.~Rudner}
 \affiliation{Department of Physics, Massachusetts Institute of Technology, Cambridge MA 02139}
\author{S.~O.~Valenzuela}
 \affiliation{MIT Francis Bitter Magnet Laboratory, Cambridge, MA 02139}
\author{K.~K.~Berggren}
 \altaffiliation[Present address: ]{EECS Department, MIT} 
 \affiliation{MIT Lincoln Laboratory, 244 Wood Street, Lexington, MA 02420}
\author{W.~D.~Oliver}
 \email{oliver@ll.mit.edu}
 \affiliation{MIT Lincoln Laboratory, 244 Wood Street, Lexington, MA 02420}
\author{L.~S.~Levitov}
 \affiliation{Department of Physics, Massachusetts Institute of Technology, Cambridge MA 02139}
\author{T.~P.~Orlando}
 \affiliation{Department of Electrical Engineering and Computer Science, Massachusetts Institute of Technology, Cambridge, MA 02139}

\date{\today}

\begin{abstract}
The energy-level structure of a quantum system plays a fundamental role in determining its behavior and manifests itself in a discrete absorption and emission spectrum~\cite{Schawlow82a}. Conventionally, spectra are probed via frequency spectroscopy~\cite{Schawlow82a,Thompson85} whereby the frequency $\nu$ of a harmonic driving field is varied to fulfill the conditions
$\Delta E = h \nu$, where the driving field is resonant with the level separation $\Delta E$ ($h$ is Planck's constant).
Although this technique has been successfully employed in a variety of physical systems, including natural~\cite{Schawlow82a,Thompson85} and artificial~\cite{Clarke88,Friedman00a,Wal00a,Berkley03a,vanderWiel03a,Hanson07a,Nakamura99a,Vion02a,Yu02a,Martinis02a,Chiorescu03a} atoms and molecules, its application is not universally straightforward, and becomes extremely challenging for frequencies in the range of 10's and 100's of gigahertz.
Here we demonstrate an alternative approach,
whereby a harmonic driving field sweeps the atom through its energy-level avoided crossings at a fixed frequency,
surmounting many of the limitations of the conventional approach.
Spectroscopic information is obtained from the amplitude dependence of the system response.
The resulting ``spectroscopy diamonds'' contain interference patterns and population inversion that serve as a fingerprint of the atom's spectrum. By analyzing these features, we determine the energy spectrum of a manifold of states with energies from 0.01 to 120 GHz$\times h$ in a superconducting artificial atom, using a driving frequency near 0.1 GHz. This approach provides a means to manipulate and characterize systems over a broad bandwidth, using only a single driving frequency that may be orders of magnitude smaller than the energy scales being probed.
\end{abstract}

\maketitle

Spectroscopy has historically been used to obtain a wide range of information on atomic and nuclear properties~\cite{Schawlow82a,Thompson85}. Early on, the determination of spectral lines helped elucidate the principles of quantum mechanics through studies of the hydrogen atom and facilitated the discovery of electron and nuclear spin.
Since then, several spectroscopy techniques to determine absolute transition frequencies (or, equivalently, wavelengths) have been developed, involving the emission, absorption, or scattering (e.g. Raman) of radiation.
The advent of tunable, coherent radiation sources at microwave and optical frequencies led to the age of modern atomic spectroscopy,
where a primary approach is to identify absorption spectra as the source frequency is varied~\cite{Schawlow82a,Thompson85}.

The study of artificial atoms at cryogenic temperatures, whose spectra extend into the microwave and millimeter wave regimes (10-300 GHz), faces different challenges.
Stable tunable microwave sources in excess of 70 GHz exist, but are expensive, and generally require multipliers which are inefficient and intrinsically noisy~\cite{Collin}.
Frequency dependent dispersion and attenuation, tight tolerances to control impedance, and multi-mode or restricted-bandwidth performance of transmission lines and waveguides~\cite{Collin}, limit the application of broadband frequency spectroscopy in these systems.
Despite these challenges, superposition states in superconducting~\cite{Friedman00a,Wal00a,Berkley03a} and semiconducting artificial atoms~\cite{vanderWiel03a} have been probed directly up to several 10's of GHz.
%
A number of leading groups have developed alternative techniques, e.g., resonant- and photon-assisted tunneling~\cite{Clarke88,Hanson07a}, which can be used to access spectroscopic information in specific systems at even higher frequencies, though each has its own advantages and limitations and may not be easily applicable to other systems.

``Amplitude spectroscopy,'' introduced in this manuscript, probes the energy level structure of a quantum system via its response to driving-field amplitude rather than frequency (Fig. 1A). It is applicable to systems with energy-level avoided crossings that can be traversed using an external control parameter. For a generic artificial atom, such longitudinal excursions throughout the energy level diagram (Fig. 1C) are realized by strongly driving the system with an external field at a fixed frequency, which may be several orders of magnitude lower than the energy-level spacing. In this limit, the system evolves adiabatically, except in the vicinity of energy-level avoided-crossings where Landau-Zener-type quantum coherent transitions occur. The quantum interference between repeated Landau-Zener transitions gives rise to interference fringes that encode information about the system's energy spectrum. By trading amplitude for frequency, the amplitude spectroscopy approach allows one to probe quantum systems with strong coupling to external fields, such as solid-state artificial atoms, over extraordinarily wide bandwidths, bypassing many of the limitations of a frequency-based approach.

\begin{figure*}
\begin{center}
\includegraphics[width=7in]{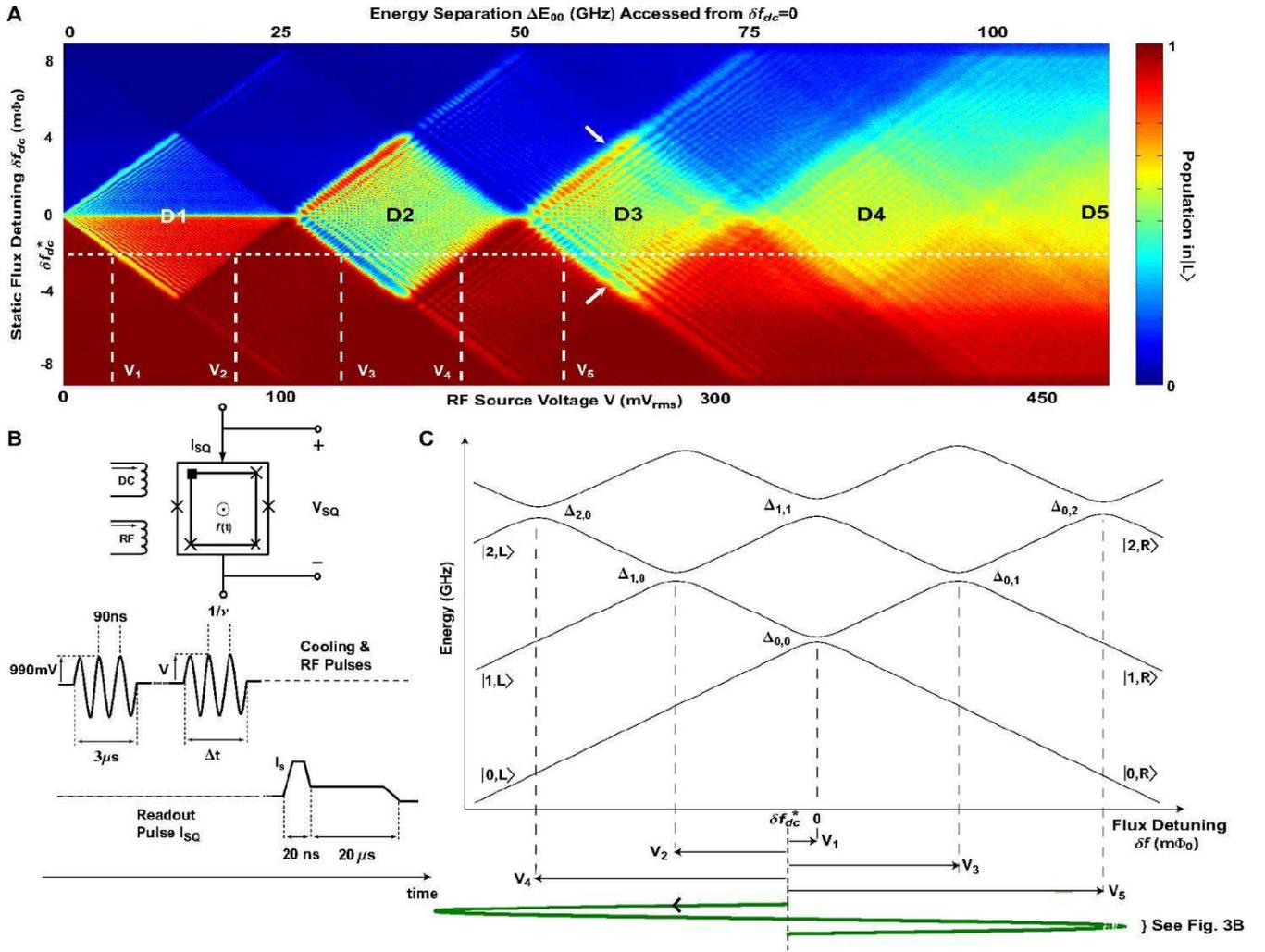}
\caption{
Amplitude spectroscopy with long-pulse driving towards saturation. (\textbf{A}) Amplitude spectroscopy diamonds. The qubit is driven at a fixed frequency $\nu=0.160$~GHz, while amplitude $V$ is swept for each static flux detuning $\delta f_{\rm{dc}}$. Color scale: net qubit population in state $|q,L\rangle$, where $L$ ($R$) labels
diabatic states of the left (right) well of the qubit double-well potential, and $q=0,1,2,...$ labels the longitudinal modes. The diamond edges
signify the driving amplitude $V$ for each value of $\delta f_{\rm{dc}}$ when an avoided level crossing is first reached (amplitudes $V_1-V_5$ for $\delta f_{\rm{dc}} = \delta f_{\rm{dc}}^*$). The main diamond regions, symmetric about $\delta f_{\rm{dc}}=0$, are labeled D1 to D5.
Arrows indicate signatures of transverse mode coupling (see Fig.~\ref{fig4}).
Top axis: the $|0,\textrm{L} \rangle - |0,\textrm{R} \rangle$ energy spacing $\Delta E_{0,0}$ accessed by $V$ from $\delta f_{\rm{dc}}=0$.
(\textbf{B})
Schematic of the qubit surrounded by a
SQUID magnetometer readout. Static and radio-frequency (RF) fields control the state of
the qubit: a 3-$\mu$s cooling-pulse followed by an amplitude-spectroscopy pulse of duration $\Delta t$ and amplitude $V$. The qubit state is determined with a synchronous readout pulse applied to the SQUID.
(\textbf{C}) Schematic energy-level diagram illustrating the relation between the driving amplitude $V$ and the avoided crossing positions for a particular static flux detuning $\delta f_{\rm{dc}}=\delta f_{\rm{dc}}^*$. The arrows represent the amplitudes $V_1-V_5$ of the RF field at which the illustrated avoided crossings are reached, marking the onset of the diamond regions in (A).}
\label{fig1}
\end{center}
\end{figure*}

We demonstrate amplitude spectroscopy with a superconducting qubit,
a solid-state artificial atom exhibiting discrete energy states~\cite{Clarke88} that can be strongly coupled to external radio frequency (RF) fields while preserving quantum coherence~\cite{Nakamura99a}.
Artificial atoms are natural systems for probing a wide range of quantum effects: coherent superpositions of macroscopic states~\cite{Friedman00a,Wal00a,Berkley03a}, Rabi oscillations~\cite{Nakamura99a,Nakamura01,Vion02a,Yu02a,Martinis02a,Chiorescu03a,Claudon04a,Plourde05a,Saito06a}, incoherent Landau-Zener transitions~\cite{Izmalkov04a}, St\"{u}ckelberg oscillations~\cite{Oliver05a,Sillanpaa06a,Berns06a,Wilson07a},
microwave cooling~\cite{Valenzuela06a,Niskanen07b,You08a},
and cavity quantum electrodynamics~\cite{Chiorescu04a,Wallraff04a,Johansson06a}
have been demonstrated with these systems.
Significant progress has also been made toward their application to quantum information science~\cite{Makhlin01a,Mooij05},
including state initialization~\cite{Valenzuela06a}, tunable~\cite{Hime06a,Ploeg07a,Niskanen07a} and long-distance~\cite{Sillanpaa07a,Majer07a} coupling, quantum control~\cite{Pashkin03a,Yamamoto03a,McDermott05a,Plantenberg07a}, quantum state tomography~\cite{Steffen06a}, and measurement~\cite{Siddiqi04a,Katz06a,Lupascu07a}.

Our qubit (Fig.~\ref{fig1}B) is a niobium superconducting loop interrupted by three Josephson
junctions (see also App.~\ref{app:supplementary_information})~\cite{Mooij99a,Orlando99a}.
The qubit potential has a two-dimensional double-well profile near flux-bias $f \approx \Phi_0/2$, parameterized by the detuning $\delta f \equiv f - \Phi_0/2$, where $\Phi_0$ is the superconducting flux quantum (Fig.~A5).
The system's energy eigenstates are comprised of transverse ($p=0,1,2,\ldots$) and longitudinal ($q=0,1,2,\ldots$) modes,
with energies controlled by 
the flux detuning $\delta f$.
%
When the potential is tilted so that resonant interwell tunneling 
is suppressed, the eigenstates closely approximate the diabatic well states localized in the left (L) and right (R) wells, 
and are associated with loop currents of opposing circulation. 
In this limit, the energies of localized states in the left (right) well increase (decrease) approximately linearly with flux detuning (Fig.~\ref{fig1}C).
Whenever the diabatic states $|p,q,\textrm{L}\rangle$ and $|p',q',\textrm{R} \rangle$ are degenerate, resonant interwell tunneling mixes them and opens avoided crossings $\Delta_{pq,p'q'}$.

We drive the qubit longitudinally with a time-dependent flux $\delta f(t) = \delta f_{\rm{dc}} - \Phi_{\rm{rf}} \sin \omega t$, with 
a harmonic term of 
amplitude $\Phi_{\rm rf}$, 
proportional to the source voltage $V$, that induces sinusoidal excursions with frequency $\nu = \omega/2\pi$ through the energy bands about a static flux bias $\delta f_{\rm{dc}}$.
Because the transverse and longitudinal modes should be decoupled for a symmetric system,
we assume initially 
that only the lowest transverse
mode is populated and use the reduced notation: $|p,q,\textrm{L}
\rangle \rightarrow |q,\textrm{L} \rangle$, $|p',q',\textrm{R}
\rangle \rightarrow |q',\textrm{R} \rangle$, and $\Delta_{pq,p'q'}
\rightarrow \Delta_{q,q'}$ (Fig.~A5B). We do observe, however,
signatures of weak excitations of transverse modes
(see Fig.~\ref{fig4} and related discussion).

\begin{figure*}
\begin{center}
\includegraphics[width=7in]{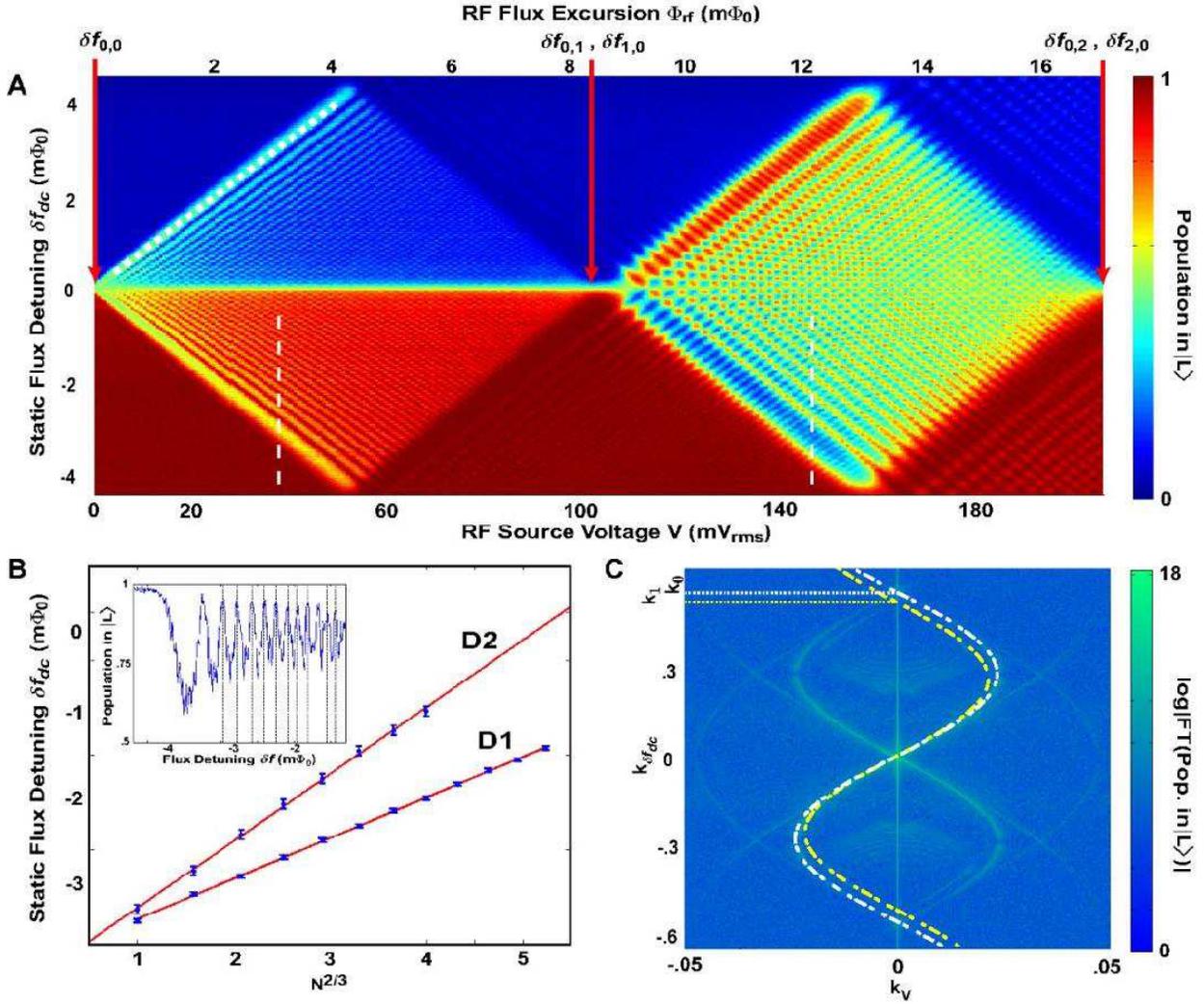}
\caption{Energy-level slopes and interference patterns. (\textbf{A}) Detail of diamonds D1 and D2 (Fig. 1A) showing interference patterns due to single (D1) and multiple (D2) avoided crossings (see text). Diamond D2 exhibits strong population inversion. Arrows mark the location of the avoided crossing positions. Energy-level slopes are extracted in (B) from the interference fringes (dashed white lines) at 43 $\textrm{mV}_{\textrm{rms}}$ (D1) and 150 $\textrm{mV}_{\textrm{rms}}$ (D2). The flux-to-voltage conversion factor is determined by the front side of D1 (dotted white line).
(\textbf{B}) Determination of the energy-level slopes for levels $|0,L\rangle$, $|0,R\rangle$, $|1,L\rangle$, and $|1,R\rangle$. Detuning location of the $N$th interference-node (see inset) vs. $N^{2/3}$ at the voltages marked with dashed lines in (A) and their corresponding linear fits (red line).
Inset: Vertical slice from diamond D1 (43 $\textrm{mV}_{\textrm{rms}}$). Interference nodes $N$
used in the main panel are indicated by dotted lines.
(\textbf{C}) 2D Fourier transform of both diamonds in (A). The sinusoids with wavenumbers $k_{0}$ and $k_{1}$ are contributions from diamonds D1 and D2, respectively, and are related to the energy-level slopes (see text).
}
\label{fig2}
\end{center}
\end{figure*}
Each experiment uses the pulse sequence shown in
Fig.~\ref{fig1}B, which consists of a harmonic cooling pulse to
initialize the qubit to its ground state~\cite{Valenzuela06a},
followed by the desired amplitude spectroscopy pulse.
The qubit state is determined with a synchronous readout pulse applied to a superconducting quantum interference device (SQUID) magnetometer (App.~\ref{app:qubit_readout}).
Using this technique, we investigate both the long and short time behavior of our
qubit, and infer the energy-level slopes $m_{q}$, and the splittings $\Delta_{q,q'}$ and 
locations $\delta f_{q,q'}$ of avoided crossings to construct the energy level diagram.

Fig.~\ref{fig1}A displays the amplitude spectroscopy of the qubit
driven towards saturation.
We choose a driving frequency $\nu$ such that, throughout the driving cycle,
$h \nu$ is generally much smaller than the instantaneous energy-level
spacing, yet the speed is large enough to make the evolution through avoided crossings
non-adiabatic.
In this regime, Landau-Zener transitions drive the system into coherent superpositions of energy eigenstates associated with different wells.
Four primary ``spectroscopy diamonds'' with large population contrast, centered about $\delta f_{\rm{dc}} = 0$ (D1, D2, D3, and D4), are observed in the data; they are flanked by eight fainter diamonds.
The diamond structures result from the
interplay between static flux detuning and driving amplitude,
which determine when the various avoided crossings are reached.
Because the onset of each diamond is associated with transitions at a particular avoided crossing, the diamond boundaries
signify the avoided crossing locations.
We use the linear relation between $V$ and $\Phi_{\rm{rf}}$ (see Fig.~\ref{fig2}A)
to obtain the avoided crossing locations $\delta f_{q,q'}$ listed in Table 1.

For a particular static flux detuning $\delta f_{\rm{dc}}=\delta f_{\rm{dc}}^* < 0$
(Figs.~\ref{fig1}A and~\ref{fig1}C) and driving
amplitude increasing from $V=0$, population transfer from
$|0,\textrm{L}\rangle$ to $|0,\textrm{R}\rangle$ first occurs when
the $\Delta_{0,0}$ avoided crossing is reached at $V = V_1$ (front
side of diamond D1, see Fig.~\ref{fig1}A).
For $V_1< V <V_2$, interferometric Landau-Zener physics~\cite{Oliver05a,Sillanpaa06a,Berns06a,Mark07a} at the $\Delta_{0,0}$ avoided crossing results in the observed fringes 
(detail, Fig.~\ref{fig2}A).
At $V = V_2$, the adjacent avoided crossing $\Delta_{1,0}$ is reached, inducing transitions between levels $|0,\textrm{R}\rangle$ and $|1,\textrm{L}\rangle$ and marking the back of D1.

For $V_2< V <V_3$, the saturated population depends on
the competition between transitions at $\Delta_{0,0}$ and $\Delta_{1,0}$,
on fast intrawell relaxation,
and to a lesser extent
on much slower interwell relaxation processes.
Because in our experiment $\Delta_{0,0} \ll h\nu \approx \Delta_{1,0}$, the result
is dominated by the dynamics at the $\Delta_{1,0}$ crossing.
Although transitions $|0,\textrm{L} \rangle \rightarrow |0,\textrm{R}\rangle$ are
induced at the $\Delta_{0,0}$ crossing, stronger transitions at $\Delta_{1,0}$
convert a substantial fraction of that population to $|1,\textrm{L}\rangle$.
This excited population quickly relaxes back to $|0,\textrm{L}\rangle$, thus suppressing the net population transfer.
For the combinations of $\delta f_{\rm{dc}}$ and $V$ where interference between successive passages through $\Delta_{1,0}$ is instead destructive, signatures of transitions due to $\Delta_{0,0}$ are visible, albeit with reduced contrast (detail, Fig.~\ref{fig2}A).

At even larger amplitudes, transitions to additional excited states become possible.
When $V>V_3$, the qubit can make transitions between $|0,\textrm{L}\rangle$ and $|1,\textrm{R}\rangle$, marking the front side of diamond D2.
The backside of this diamond is marked by the amplitude $V=V_4$ that reaches $\Delta_{2,0}$, allowing transitions between $|0,\textrm{R}\rangle$ and $|2,\textrm{L}\rangle$.
This description can be extended straightforwardly to the remainder of the spectrum. 
In this device, we did not find explicit signatures of coherent multi-path traversal between the $\delta f < 0$ and $\delta f > 0$ regions of the energy-level diagram (e.g., via avoided crossings
$\Delta_{1,1}$ and $\Delta_{2,2}$, $\ldots$).
%
\begin{center}
\begin{figure*}[ht]
\includegraphics[width=7in]{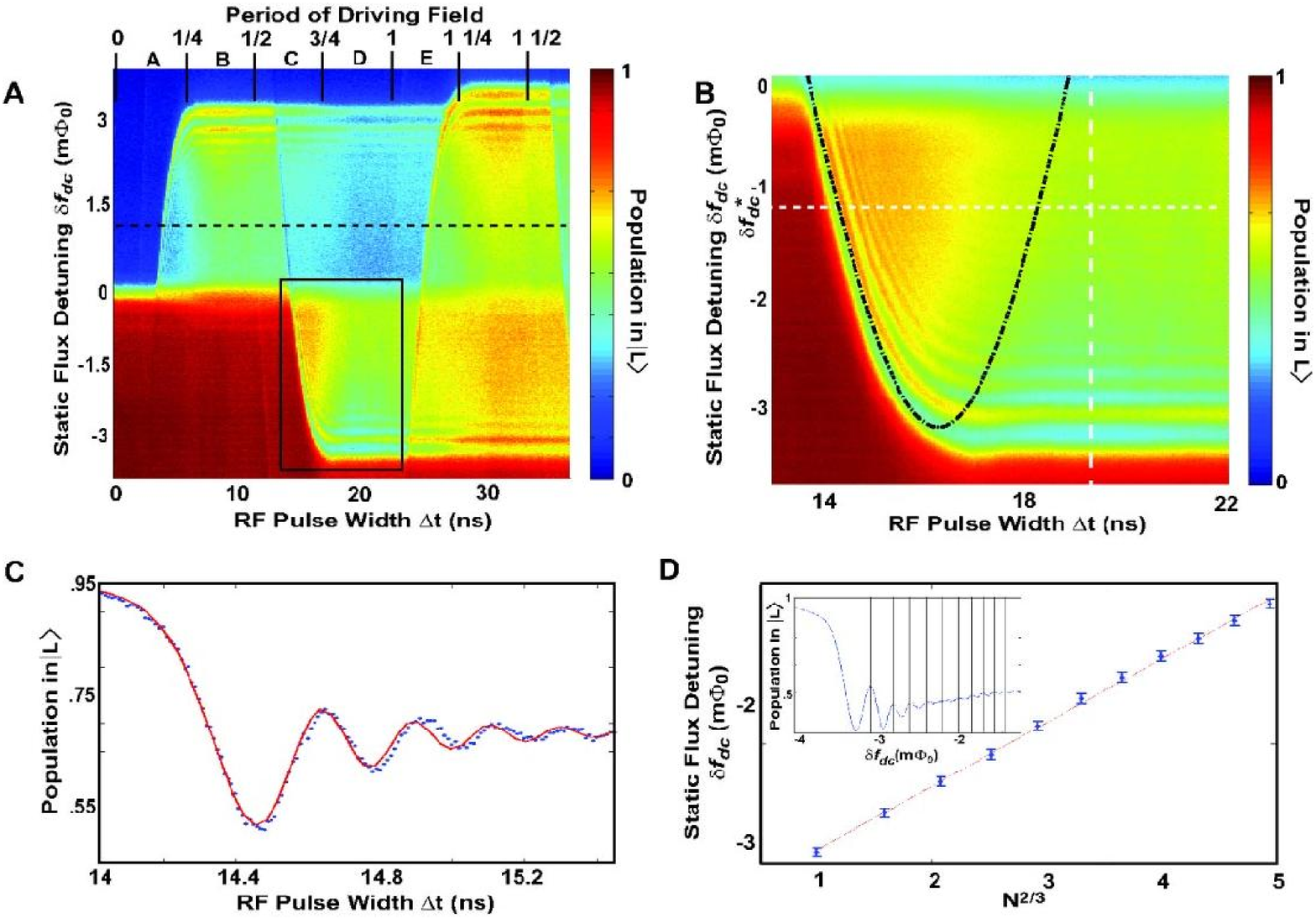}
\caption{Amplitude spectroscopy with short-pulse driving. (\textbf{A}) Qubit response to a short RF pulse of variable length $\Delta t$ as a function of static flux detuning $\delta f_{\rm{dc}}$, with $V=181 \textrm{ mV}_{\textrm{rms}}; \nu = 0.045$ GHz.  Top axis: driving field period (regions A-E); the maximum pulse width corresponds to $\sim$1.5 oscillation periods. The scan is performed at an amplitude value in the left side of diamond D3 (Fig.A11), which reaches all crossings through $\Delta_{0,2}$ and $\Delta_{2,0}$.
(\textbf{B}) Detail of the interference pattern in the boxed region of (A). The black curve marks the pulse width $\Delta t$ at which the sinusoidal flux-excursion first exceeds and, subsequently, returns through $\Delta_{0,2}$ at flux detunings $\delta f_{\rm{dc}}$.
(\textbf{C}) Temporal oscillations along the horizontal line in (B) at the specific static flux bias $\delta f_{\rm{dc}} = \delta f_{\rm{dc}}^*$, fitted by a Landau-Zener model with damping (red line, see text). (\textbf{D}) $N$th interference node versus $N^{2/3}$ along the vertical line in (B) and best linear fit. Inset: interference pattern along vertical line in (B) and node locations.
}
\label{fig3}
\end{figure*}
\end{center}

There are several remarkable features associated with the amplitude spectroscopy shown in Fig.~\ref{fig1}A.
First, we are able to probe the qubit continuously over extraordinarily wide bandwidth using a single driving frequency of only 0.16 GHz.
The highest diamond (D5) in Fig.~\ref{fig1}A results from transitions to energy levels
more than 100 GHz$\times h$ above the ground state.
Even at such high energy levels, our artificial atom retains its energy-level structure in the presence of the strong driving used to access them.

Second, diamond D2 exhibits strong population inversion due to competition between transitions at avoided crossings $\Delta_{0,1}$ and $\Delta_{1,0}$ combined with fast intrawell relaxation to
$|0,\textrm{L} \rangle$ and $|0,\textrm{R} \rangle$ (Fig.~\ref{fig2}A).
The transition rates at $\Delta_{0,1}$ and $\Delta_{1,0}$ exhibit strong oscillatory behavior due to Landau-Zener interference, constructive or destructive, depending on the values of $\delta f_{\rm{dc}}$ and $V$.
The competition between these rates leads to a checkerboard pattern symmetric about $\delta f_{\rm{dc}}=0$ with alternating regions of strong population inversion and efficient cooling.
Similar checkerboard patterns are present in the higher diamonds.
The population inversion observed here is incoherent, and can serve as the pump for a single-atom laser~\cite{Astafiev07a}.

The energy-level separation $\Delta E_{q,q'} \equiv h (\vert m_{q}\vert + \vert m_{q'}\vert) (\delta f_{\rm{dc}} - \delta f_{q,q'})$ between states $|q,\textrm{L}\rangle$ and $|q'\textrm{,R} \rangle$ is proportional to the net flux detuning from the location $\delta f_{q,q'}$ of the avoided crossing  $\Delta_{q,q'}$,
and to the sum of the magnitudes of the energy-level slopes $m_{q}$ and $m_{q'}$.
Because the relative phase accumulated between the $|q,\textrm{L}\rangle$ and $|q',\textrm{R}\rangle$ components of the wave function over repeated Landau-Zener transitions is sensitive to $\Delta E_{q,q'}$, the slopes can be derived from the interference patterns which arise when varying $\delta f_{\rm{dc}}$.
The $N$th node in the interference pattern occurs when a relative phase of $2\pi N$ is picked up during the qubit's excursion beyond the avoided crossing. The detuning locations of the nodes (inset, Fig.~\ref{fig2}B) follow the power-law $s_{qq'}N^{2/3}$ with a prefactor $s_{qq'}$ related to the energy-level slopes (App.~\ref{app:slope_extraction}):
\begin{equation}
 \label{slope_from_Fig2}
    \vert m_{q}\vert + \vert m_{q'}\vert =
        a
        \nu \sqrt{ \alpha V / s_{qq'}^{3}}
        \textrm{  } \left( \frac{\textrm{GHz}}{\textrm{m}\Phi_0} \right),
\end{equation}
where $a = 3 \pi / 2 \sqrt{2}$. The factor $\alpha$ is the frequency-dependent conversion factor between RF flux and source voltage; 
its value $\alpha=0.082 \textrm{ m}\Phi_0/\textrm{mV}_{\textrm{rms}}$ is inferred from the slope
of the front edge of the first diamond (Fig.~\ref{fig2}A).
Fig.~\ref{fig2}B shows the
$N^{2/3}$ power-law fits to the nodes of the
vertical slices in diamonds D1 and D2 which are used
to extract $m_{0}$ and $m_{1}$ (Fig.~\ref{fig2}A, dashed vertical lines),
where we take $\vert m_{q}\vert=\vert m_{q'}\vert \equiv m_{q}$ for $q=q'$ in our system.
The slopes are obtained sequentially from the fitted values $s_{qq'}$ in
Eq.~\ref{slope_from_Fig2}, starting with $2m_{0} = 2.88$ GHz/m$\Phi_0$, followed by
$m_{0} + m_{1} = 2.534$ GHz/m$\Phi_0$; their values are
summarized in Table 1.

The relation between the slopes $m_0$ and $m_1$ is most clearly exhibited by the 2D Fourier transform of the diamonds in Fig.~\ref{fig2}C (App.~\ref{app:2DFT}).
The observed structure in the first two diamonds
arises from the underlying ``Bessel-function staircases'' 
of multi-photon resonances 
associated with transitions between the lowest four energy levels,
where the $n$-photon absorption rate depends on driving amplitude through the square of the $N$th-order Bessel function~\cite{Oliver05a,Berns06a,Wilson07a}.
Using Fourier analysis, the apparently complicated mesh of overlapping Bessel functions is transformed to a pair of sinusoids with periodicity related to the energy level slopes,
$k_V = \pm \alpha g \sin \left(k_{\delta f_{\rm{dc}}}/g\right)$, where
$g=2(|m_{q}|+|m_{q'}|)/\nu$~\cite{Rudner08a}.
The sinusoid associated with $q = q' = 0$ arises from the transitions at
$\Delta_{0,0}$, while the second sinusoid with 
$q =0,\, q' = 1$ and $q = 1, \, q' = 0$ is degenerate and arises from the transitions at
$\Delta_{0, 1}$ and $\Delta_{1, 0}$.  All four diamonds and their
individual Fourier transforms are presented in Figs.~A6-A10.
\begin{table}
 \begin{center}
  \begin{tabular}{| c | c | c | c |}
    \hline
    Crossing    &   Location     &   Coupling    &   Slope   \\
    $\textrm{q},\textrm{q}'$     &   $\delta f_{\textrm{q},\textrm{q}'} (\textrm{m}\Phi_0)$ &  $\Delta_{\textrm{q},\textrm{q}'}/h$ (GHz) & $\textrm{m}_{\textrm{q}'}$ (GHz/m$\Phi_0$) \\ \hline
    $0,0$   &   $0$              &   $0.013 \pm 0.001$  &   $1.44 \pm 0.01$ \\ \hline
    $0,1$   &   $8.4 \pm 0.2$    &   $0.090 \pm 0.005$  &   $1.09 \pm 0.03$ \\ \hline
    $0,2$   &   $17.0 \pm 0.2$   &   $0.40 \pm 0.01$    &   $0.75 \pm 0.04$ \\ \hline
    $0,3$   &   $25.8 \pm 0.4$   &   $2.2 \pm 0.1$      &   $0.49 \pm 0.08$ \\ \hline
  \end{tabular}
  \caption{Energy spectrum parameters of a superconducting artificial atom determined using amplitude spectroscopy}
 \end{center}
\end{table}

Valuable additional information about the energy level spectrum and temporal coherence is gained by performing amplitude spectroscopy over short time scales (Fig.~\ref{fig3}).
Temporal-response measurements are performed by initializing the system to the ground state at detuning $\delta f_{\rm{dc}}$ and then applying an RF field pulse of a variable length $\Delta t$, with fixed frequency and amplitude.
The phase of the sinusoid is carefully adjusted to maintain the timing and directionality of the RF-flux excursion through the energy levels between pulses.
When the pulse ends abruptly at time $\Delta t$,
the state of the system 
is preserved as the flux detuning instantaneously returns to $\delta f_{\rm{dc}}$ (finite decay-time corrections are discussed below).

\begin{figure*}[t]
\includegraphics[width=6.5in]{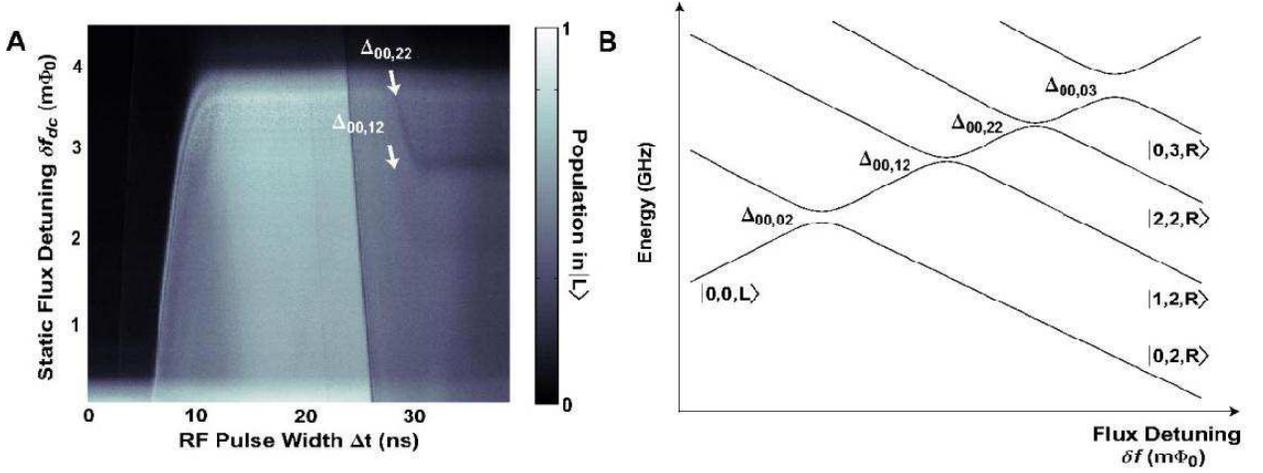}
\caption{Identification of transverse states of the qubit's double-well potential. (\textbf{A}) Qubit response to a short RF pulse of variable length $\Delta t$ as a function of static flux detuning $\delta f_{\rm{dc}}$, with $V=179 \textrm{ mV}_{\textrm{rms}}; \nu = 0.025$ MHz. The scan is performed at an amplitude value in the right side of diamond D3 (Fig.~A12), where the crossing $\Delta_{03,00}$ (but not $\Delta_{00,03}$) is reached. The signatures of two crossings with transverse states, $\Delta_{12,00}$ and $\Delta_{22,00}$, are indicated. (\textbf{B}) Schematic energy-level diagram showing the locations of the transverse states.}
\label{fig4}
\end{figure*}

This is exemplified in Fig.~\ref{fig3}, where parameters are tuned to investigate the
$\Delta_{2,0}$ level crossing (Fig.~A11).
For example, starting in the ground state at positive detuning ($\delta f_{\rm{dc}} > 0$),
the qubit is driven through $\delta f(t) < 0$, diabatically crossing $\Delta_{0,0}$ and $\Delta_{1,0}$ at the beginning of the first quarter-period with no significant population transfer (beginning of region A).
Significant population transfer first occurs in region A when $\Delta_{2,0}$ is 
reached.
The 
onset of population transfer is followed by brief temporal oscillations, which we use to find the splitting $\Delta_{2,0}$ (see below). The population becomes stationary after the qubit returns 
through $\Delta_{2,0}$ in the second quarter period (region B).

In the third quarter-period (region C), the driving pulse takes the qubit through the excited state avoided crossing $\Delta_{0,2}$ located on the opposite side of the level diagram (positive flux bias, Fig.~\ref{fig1}C).
This event is marked by a second sharp population transfer. 
The population subsequently remains nearly constant 
(region D) until a third abrupt population transfer occurs during the first quarter of the second period (region E), which signals the qubit's return through $\Delta_{2,0}$ and the repetition of the driving cycle.
The population transfer does not reach its furthest extent in flux during the first half-period (as it does for the subsequent half-periods) because our pulse shape has slightly lower amplitude for times smaller than 5 ns.

The observed response over short time scales is not 
symmetric about $\delta f_{\rm{dc}}=0$.
When starting in the ground state at static bias $\delta f_{\rm{dc}} = \delta f_{\rm{dc}}^* < 0$, the system is drawn deep into the ground state during the first half-period, without inducing any transitions (see Fig~\ref{fig1}C).
It is only during the second half-period 
that $\Delta_{0,2}$ is finally reached and significant population transfer occurs.
The detailed time dependence of population in this interval is shown in Fig.~\ref{fig3}B. 
The curved line
in Fig.~\ref{fig3}B marks the pulse width $\Delta t$ at
which the sinusoidal flux excursion first exceeds and,
subsequently, returns through $\Delta_{0,2}$ for different flux
detunings $\delta f_{\rm{dc}}$; the sinusoidal excursion about the specific static flux
$\delta f_{\rm{dc}}^*$ is correspondingly indicated in Fig.~\ref{fig1}C.

The temporal oscillations displayed in Fig.~\ref{fig3}B and Fig.~\ref{fig3}C can be understood qualitatively as a Larmor-type precession, or ``ringing,'' that results after the qubit is swept through the avoided crossing. 
In a pseudo-spin-1/2 picture where the qubit
states are identified with up and down spin states relative to a fictitious z-axis,
the qubit precesses about a tipped effective magnetic field which steadily increases in magnitude and rotates toward the z-axis. 
This picture 
is consistent with a temporal analysis of the standard Landau-Zener problem, in which a linear ramp with velocity $v$ sweeps the qubit 
through the avoided crossing. 
In the perturbative (non-adiabatic) limit~(\ref{app:Fresnel}), this model yields the transition probability
\begin{equation}
 \label{Fresnel_equation}
 P(t) = \frac{\Delta_{0,2}^2}{4} \left| \int_{t_0}^t e^{iv t'^2/2}dt' \right|^2.
\end{equation}
The integral in Eq.~\ref{Fresnel_equation} often arises in the context of optical diffraction, giving rise to Fresnel oscillations 
similar to the coherent oscillations observed in Fig.~\ref{fig3}C.

Although Eq.~\ref{Fresnel_equation} captures the essential features of the data in Fig.~\ref{fig3}C, to obtain a quantitative fit we must account for the non-abrupt ending of the pulse.
This transient slightly modifies the total precession phase accumulated before read-out, and
adds a small Mach-Zehnder-type interference due to the finite ramp speed back through the avoided crossing $\Delta_{0,2}$.
We found remarkable agreement between the data and a simulation of the Bloch dynamics of the two level system near $\Delta_{0, 2}$ which included longitudinal sinusoidal driving up to time $t = \Delta t$ followed by a rapid 
decay over approximately 2 ns, and intrawell relaxation with a rate of 0.65 ns$^{-1}$
(Fig.~\ref{fig3}C).
The value of $\Delta_{0,2}$ is extracted as a fitting parameter 
and, in this regime, is largely insensitive to the details of the pulse decay and intrawell relaxation.

As in the case of the long-time data, the energy-level slopes can be extracted
from the vertical fringes (Fig.~\ref{fig3}B) using the $N^{2/3}$ power-law fitting (Fig.~\ref{fig3}D) and Eq.~\ref{slope_from_Fig2}.
We used Eq.~\ref{slope_from_Fig2} to infer $m_{2}$ and $m_{3}$ from the sums $m_{0} + m_{2} = 2.189$ GHz/m$\Phi_0$ and $m_{0} + m_{3} = 1.929$ GHz/m$\Phi_0$.
The short-time amplitude spectroscopy procedure was applied to
obtain $\Delta_{q,q'}$ for diamonds D2-D4 and slopes $m_{q}$ for diamonds D3-D4, and they
are presented in Table 1 ($\Delta_{0,0}$ was previously obtained using the rate-equation approach~\cite{Berns06a}).

So far we have focused only on the strongly coupled longitudinal modes.
However, the lack of perfect symmetry allows us to probe excited transverse modes within our driving scheme as well.
The population transfer is relatively weak,
indicating small deviations from an ideally symmetric double-well
potential and longitudinal driving. Signatures of these states
appear in diamonds D3 and D4 (e.g., arrows in Fig.~\ref{fig1}A and Fig.~A13). The temporal
response to a short RF pulse taken on the
front side of diamond D3 (Fig.~A12) is shown in Fig.~\ref{fig4}A.
The front side of diamond D3 results from accessing $\Delta_{02,00}$ during the first half-period, where some population is transferred from the right to the left wells (Fig.~\ref{fig4}A), and we have used the full notation explicitly indicating both longitudinal and transverse modes.
Two weak population transfers are identified during the second half-period between the known positions of the longitudinal avoided crossings $\Delta_{00,02}$ and $\Delta_{00,03}$. This result is in agreement with simulations of the qubit Hamiltonian~\cite{Orlando99a}, which indicate that two transverse modes $|1,2,\textrm{R} \rangle$ and $|2,2,\textrm{R} \rangle$ exist in this region, as illustrated in Fig.~\ref{fig4}B.
Although we can identify their locations, the values of $\Delta_{00,12}$ and $\Delta_{00,22}$ are not conclusively determined from this measurement, because the fringe contrast of their temporal oscillations are small compared with those of the adjacent longitudinal crossings $\Delta_{00,02}$ and $\Delta_{00,03}$.

The amplitude spectroscopy technique demonstrated here is generally applicable to systems with
traversable avoided crossings, including both artificial and natural atomic systems. Due to the sensitivity of interference conditions and transition probabilities to system parameters, it is a useful tool to study and manipulate quantum systems. The technique is extensible to anharmonic excitation,
e.g., one can utilize arbitrary-waveform excursions through the energy levels and targeted harmonic excitations to achieve desired transitions. This type of hybrid driving has been very recently demonstrated with Cs~\cite{Mark07b} and Rb~\cite{Lang08a} atoms about Feshbach resonances, systems containing weakly-coupled levels that are otherwise 
challenging to address with a standard frequency-based approach.
The amplitude spectroscopy technique is complementary to frequency spectroscopy, allowing the characterization of a quantum system, and, more generally, should open new pathways for quantum control~\cite{Ashab07a}.

We thank A. Shytov, J. Bylander, B. Turek, A.J. Kerman, and J. Sage for helpful discussions;
V. Bolkhovsky, G. Fitch, E. Macedo, P. Murphy, K. Parrillo, R. Slattery, and T. Weir at MIT Lincoln Laboratory for technical assistance. This work was supported by AFOSR and LPS (F49620-01-1-0457) under the DURINT program,
and by the U.S. Government. The work at Lincoln Laboratory was sponsored by the US DoD under Air Force Contract No. FA8721-05-C-0002.

\appendix
\section{Supplementary Information}
\label{app:supplementary_information}
\subsection{Device Fabrication and Parameters}
\label{app:dev_fab}
The device was fabricated at MIT Lincoln Laboratory.
It has a critical current density $J_{\rm c}
\approx 160 \,{\rm A/cm^2}$, and the characteristic Josephson and
charging energies are $E_{\rm J} \approx (2\pi\hbar)300\,{\rm
GHz}$ \,\,{\rm and}\,\, $E_{\rm C} \approx (2\pi\hbar)0.65\,{\rm
GHz}$ respectively. The ratio of the qubit JJ areas is $\alpha
\approx 0.84$, and $\Delta\approx (2\pi\hbar)10\,{\rm MHz}$.  The
qubit loop area is 16 $\times$ 16 $\mu$m$^2$, and its self
inductance is $L_{\rm{q}} \approx 30$ pH. The SQUID Josephson
junctions each have critical current $I_{\rm{c0}} \approx 2$
$\mu$A. The SQUID loop area is 20 $\times$ 20 $\mu$m$^2$, and its
self inductance is $L_{\rm{S}} \approx 30$ pH. The SQUID junctions
were shunted with 2 1-pF on-chip capacitors.  The mutual coupling
between the qubit and the SQUID is $M \approx 25$ pH.
\subsection{Potential Energy of the PC Qubit}
\label{app:PE_PC_qubit}
The potential energy of the PC Qubit is a 2D anisotropic periodic
potential with double-well structures at each lattice site.  After
designing for negligible inter-lattice-site tunneling, the
potential energy can be visualized as a single double-well, as
seen in Fig.~A5a. It is convenient to parameterize the
potential energy U with phase variables
$\varphi_m = (\varphi_1-\varphi_2)/2$ and
$\varphi_p = (\varphi_1+\varphi_2)/2$, where $\varphi_1$ and $\varphi_2$
are the phases across junctions 1 and 2, respectively, in Fig.1B.
It is also convenient to plot $\textrm{U}/ \textrm{E}_\textrm{j}$, where $\textrm{E}_\textrm{j}=
\Phi_0 \textrm{I}_\textrm{c}/2 \pi$ and $I_c$ is the critical current of the larger junction.
When the double-well potential is
symmetric and the qubit is driven symmetrically, one can reduce the
2D potential to a 1D double-well along the $\varphi_m$ direction, as seen in
Fig.~A5b. This is the longitudinal direction along which the
qubit circulating current varies through the phase $\phi_m$~\cite{Orlando99a}.
\begin{figure}[h]
 \label{OSMfig1}
 \center{
 \includegraphics[width=3in]{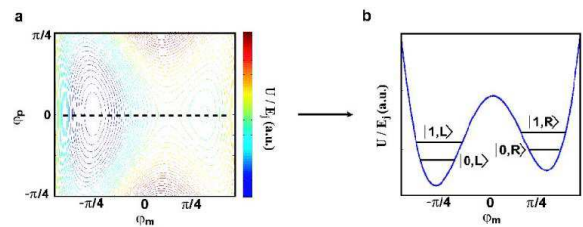}
 \caption{\textbf{2D and 1D double-well potential energies (see text). a,}
 Contour plot of 2D double-well potential energy for our qubit at
 $\delta f_{0} = 0.46 \Phi_0$, far detuned from
 the symmetry point $\delta f_{\rm{dc}}=0$. When the transverse
 quantum modes can be ignored, the potential energy can be treated
 as a 1D double-well along the dashed line pictured. \textbf{b,}
 1D double-well potential energy for $\delta f_{0} = 0.495 \Phi_0$.
 The wells are slightly tipped to the left and the four lowest energy eigenstates
 are shown.}}
\end{figure}

\subsection{Qubit Readout}
\label{app:qubit_readout}
The qubit states are read out using a DC-SQUID, a sensitive
magnetometer that distinguishes the flux generated by the qubit
persistent currents, $I_{\rm q}$. The readout is performed by
driving the SQUID with a 20-ns ``sample'' current $I_{\rm{s}}$
followed by a 20-$\mu$s ``hold'' current (Fig.~1B). The SQUID will
switch to its normal state voltage $V_{\rm{s}}$ if $I_{\rm{s}} >
I_{\rm{sw,L}}$ ($I_{\rm{s}}
> I_{\rm{sw,R}}$), when the qubit is in state $|L\rangle$ ($|R\rangle$).
By sweeping the SQUID current $I_{\rm{s}}$ and qubit flux detuning $\delta f_0$, while monitoring the
presence of a SQUID voltage over many trials, we generate a
cumulative switching-current distribution function. From this distribution,
we extract a best-estimator line in the space of $I_{\rm{s}}$ and $\delta f_0$
that allows us to characterize the population of state $|L\rangle$ for a given flux detuning.

\subsection{Implementation}
\label{app:implementation}
The experiments were performed in a dilution refrigerator at a
temperature of approximately 20 mK. The device was magnetically shielded with
4 Cryoperm-10 cylinders and a superconducting enclosure. All
electrical leads were attenuated and/or filtered to minimize
noise. The electrical temperature of the device in the absence of microwave cooling
was approximately 40 mK. After applying the microwave cooling pulse (Fig.1B), the effective
temperature of the qubit was less than 3 mK. Microwave cooling enabled the data to be obtained at a repetition rate of 10 kHz, much faster than the intrinsic equilibration rate due to interwell relaxation.
For all experiments, the static flux detuning was swept in
6-$\mu\Phi_{0}$ steps, and the RF amplitude was scanned in 0.5-mV steps (at the source).
The pulse width was scanned in steps of 0.005 ns to .1 ns, and each data point
comprised an average of 500 to 30,000 trials, depending on the desired resolution.

\section{Supplementary Discussion}
\label{app:supplementary_discussion}
\subsection{Slope Extraction from Landau-Zener Interference Patterns}
\label{app:slope_extraction}
The interference between sequential Landau-Zener transitions at an isolated avoided crossing is sensitive to the relative phase
\begin{eqnarray}
    \phi = 2\pi \int_{t_1}^{t_2} \Delta E(t')\,dt'
\end{eqnarray}
accumulated by the two components of the wave function between the first and second traversals of the avoided crossing.
Here $\Delta E(t')$ is the instantaneous diabatic energy level separation at time $t'$, and $t_{1,2}$ are the times of the first and second traversals, respectively. We note that the energy $\Delta E$ is measured in frequency units (GHz), and this is why the expression contains the factor $2 \pi$ rather than $1 / \hbar$.

For demonstration, we focus on the interference in the first diamond where the driving $\delta f(t) = -\delta f_{\rm{dc}} + \alpha V \cos \omega t$ sweeps the qubit through only the lowest avoided crossing $\Delta_{0,0}$.
Using the definition of the energy level slopes given in the text, this gives
\begin{eqnarray}
  \Delta E(t) \approx 2 \vert m_0 \vert (\alpha V - \delta f_{\rm{dc}}) - \vert m_0 \vert \alpha V\, \omega^2 t^2,
\end{eqnarray}
where we have fit the driving $\delta f(t)$ near the maximum of $\cos \omega t$ to a parabola.

By setting $\Delta E(t_*) = 0$, we find the initial and final crossing times $t_{1,2} = \mp t_*$, with $t_* \equiv \sqrt{\frac{2(\alpha V  - \delta f_{\rm{dc}})}{\alpha V \omega^2}}$.
In the parabolic approximation to the driving signal, the phase accumulated between crossings is
\begin{eqnarray}
  \phi &=& 2 \pi \int_{-t_*}^{t_*} \left(2 \vert m_0 \vert (\alpha V - \delta f_{\rm{dc}}) - \vert m_0 \vert \alpha V \, \omega^2 t'^2\right) dt' \\
      &=& 2\vert m_0 \vert \, \frac{8 \pi}{3}\, (\alpha V - \delta f_{\rm{dc}}) \, t_*.
\end{eqnarray}

Using the quantization condition for interference, $\phi = 2 \pi N$, and the definition of $t_*$, we find the values of static flux detuning $\{\delta f_{\rm{dc}}^{(N)}\}$ where interference occurs with driving source voltage $V$:
\begin{eqnarray}
\label{quantCond}
  2\pi N = 2 \vert m_0 \vert \, \frac{8 \pi \sqrt{2}}{3} \, \frac{(\alpha V - \delta f_{\rm{dc}}^{(N)})^{3/2}}{(\alpha V)^{1/2} \omega}.
\end{eqnarray}
Rearranging Eq. (\ref{quantCond}) and using $\omega = 2 \pi \nu$ we find
\begin{eqnarray}
    \label{detuningN}
    \delta f_{\rm{dc}}^{(N)} = \alpha V - \left(\frac{3 \pi \nu}{2 \sqrt{2}} \frac{\sqrt{\alpha V}}{2\vert m_0\vert}\right)^{2/3} N^{2/3},
\end{eqnarray}
where the prefactor to $N^{2/3}$ is identified with the fit parameter $s_{0,0}$ in the text.

Expression (\ref{detuningN}) can be generalized to any avoided crossing $\Delta_{q,q'}$ by making the replacement $2 \vert m_0\vert \rightarrow \vert m_q\vert + \vert m_{q'}\vert$, from which we arrive at Eq. (1) in the text.

\subsection{2D Fourier Transform of Amplitude Spectroscopy Diamonds}
\label{app:2DFT}
The amplitude spectroscopy plots in Figs.~1 and 2 of the main text
display structure on several scales.
On the largest scale, the boundaries of the ``spectroscopy diamonds'' 
are readily identifiable.
The interiors of the diamonds are textured by fringes arising from the interference between successive Landau-Zener transitions at a single or multiple avoided crossings.
On an even smaller scale, these fringes are composed of a series
of horizontal multiphoton resonance lines.
In order to extract information from these small scale structures,
it is helpful to apply a transformation that is able to invert
length scales; the two-dimensional Fourier transform (2DFT)
provides this service.

An analytic treatment is presented in detail in Ref.~\cite{Rudner08a}; the main conclusions are presented here. The treatment in Ref.~\cite{Rudner08a} is applicable to the perturbative driving regime $\Delta^2/\hbar v \ll 1$,
where $\Delta$ is
the splitting at the largest traversed avoided crossing and
$v=dE/dt$ is the speed of sweeping the qubit through level crossing.
In our device
for the driving frequencies employed, this condition is satisfied
in the first two diamonds, where we find good agreement between
the analytical treatment, numerics, and the data. For higher
diamonds where the dynamics are non-perturbative, more complicated
behavior is observed. The numerical approach can still be
employed, but in practice
we find that in such cases it is more efficient to extract the
desired information directly from the short-time dynamics as
described in the discussion of Fig. 3 in the text.

The internal structure of the first diamond arises from repeated passages through a single weak avoided
crossing.
As discussed in~\cite{Rudner08a}, the primary features of the 2D Fourier Transform of qubit magnetization for this perturbative driving regime can be understood by studying the 2DFT of the transition rate at the qubit level crossing.
The 2DFT of the transition rate displays
intensity concentrated along the curve
\begin{eqnarray}\label{eq:sinusoid_dimensionless}
  k_{\tilde V} = \pm \frac{2}{\nu} \sin \left(\frac{\nu}{2}\,k_{\delta \tilde f_{\rm{dc}}}\right),
\end{eqnarray}
where the flux detuning and the driving signal are measured in the energy units:
$\delta \tilde f_{\rm{dc}}=2|m_0| \delta f_{\rm{dc}}$ and $\tilde V = 2|m_0|\alpha V$.
After going back to the physical units, Eq.(\ref{eq:sinusoid_dimensionless}) gives
\begin{eqnarray}\label{eq:sinusoid_2m_0}
  k_V = \pm \frac{4|m_0|\alpha}{\nu} \sin \left(\frac{\nu}{4|m_0|}\,k_{\delta f_{\rm{dc}}}\right),
\end{eqnarray}
The simplicity of the result can be
traced to the fact that the curve (\ref{eq:sinusoid_dimensionless}) reproduces
time evolution of the quantum phase of the qubit~\cite{Rudner08a} which is harmonic for harmonic driving.

Most strikingly, the apparently distinct phenomena of interference
fringes and multiphoton resonances observed in the real space
image are manifested as a single smooth curve in Fourier space.
This structure can be understood by considering $k_V$ and
$k_{\delta f_{\rm{dc}}}$ to be smoothly varying functions of the spatial
coordinates $(V, \delta f_{\rm{dc}})$. Through a stationary phase
analysis of the Fourier integrals, one finds the mapping between
real space patches and regions of $k$-space depicted in
Fig.~A10a.

In numerical simulations we found that the steady-state
magnetization in the second diamond was well reproduced by a
simple rate model based on incoherently adding two additional
transition rates to account
for transitions at the avoided crossings with the left and right
first excited states. These additional rates are calculated using
values of $\Delta$ appropriate for the excited state avoided
crossings (approximately 90 MHz), and also take into account the
different slopes of the ground and excited state energy levels.

In the above model, the Fourier image of diamond two is
the sum of the Fourier transforms of the three relevant
transition rates. Due to the difference in the dispersion of the lowest and second lowest qubit energy levels versus dc flux bias,
for the $\Delta_{01}$ and $\Delta_{10}$ level crossings we have $\delta \tilde f_{\rm{dc}}=(|m_0|+|m_1|) \delta f_{\rm{dc}}$,
$\tilde V=(|m_0|+|m_1|)\alpha V$.
This gives a sinusoid
\begin{eqnarray}\label{eq:sinusoid_m_0+m_1}
  k_V = \pm \frac{2(|m_0|+|m_1|)\alpha}{\nu} \sin \left(\frac{\nu}{2(|m_0|+|m_1|)}\,k_{\delta f_{\rm{dc}}}\right)
\end{eqnarray}
with the wavelength different from that of (\ref{eq:sinusoid_2m_0})
by the ratio of the slopes $2|m_0|/(|m_0|+|m_1|)$.
Both sinusoids can be seen in the Fourier transform of the second diamond, shown in Fig.~A7, which indicates that the transitions at the $\Delta_{00}$ crossing and the $\Delta_{01}$ ($\Delta_{01}$) crossings contribute to the observed pattern. From the measured ratio of the sinusoids' wavelengths we obtain the ratio of the slopes of the qubit energy levels $m_0/m_1$ without any fitting parameters.

In addition, in the Fourier transform of the second diamond the real-space integration samples a more limited sector of the fringes arising from each of
the avoided crossings as compared to the case of the first diamond (see
Fig.~A10b). As a result, the regions around the middle part of a sinusoid,
$\frac{\nu}{2(\vert m_1 \vert + \vert m_{0} \vert)} k_{\delta f_{\rm{dc}}}  \approx (n + 1/2)\pi$,
are absent from the sinusoids
arising from the $\Delta_{01}$ and $\Delta_{10}$ crossings, while the regions around the nodes,
$\frac{\nu}{4|m_0|} k_{\delta f_{\rm{dc}}} \approx n\pi$,
are absent from the
sinusoid arising from transitions at the $\Delta_{00}$ crossing.

\subsection{Perturbative Treatment of Fresnel Oscillations}
\label{app:Fresnel}

The time-dependent oscillations observed in temporal-response measurements (see Fig. 3C) result from Larmor precession about a tilted axis following the qubit's transit through an avoided crossing.
In the regime where the Landau-Zener transition probability is small, we use a perturbative model to relate these oscillations to the well-known Fresnel integral. 

By linearizing the sinusoidal driving signal near the moment of traversal through the avoided crossing, $t_*$, we arrive at the familiar Landau-Zener Hamiltonian
$\hat{H}(t) = (\hbar / 2) (vt \, \hat{\sigma}^z + \Delta \hat{\sigma}^x),$
where $v = h(\vert m_q\vert + \vert m_{q'}\vert) \Phi_{\rm RF}\cos\omega t_*$ is the sweep velocity and $\Delta$ is the splitting at the avoided crossing.
Next, we transform to a non-uniformly rotating frame by
$\Ket{\psi_R (t)} = e^{i \phi(t) \hat{\sigma}^z}\Ket{\psi (t)},$
where $\phi(t) \equiv -\frac{1}{2}\int_0^t vt'\, dt' = -\frac{1}{4} v t^2$.
The rotating frame Hamiltonian is purely off diagonal, with $\Delta_R(t) \equiv \Delta \exp(-ivt^2/2)$,
\begin{eqnarray}
  \hat{H}_R = \frac{\hbar}{2}
  \left(
    \begin{array}{cc}
      0 & \Delta_R(t) \\
      \Delta_R^*(t) & 0.
    \end{array}
    \right),
\end{eqnarray}

We now 
expand the system's time evolution operator $\hat{U}(t, t_0)$ to first order in $\Delta$:
\begin{eqnarray}
  \hat{U}(t, t_0) = \hat{1} - \frac{i}{\hbar} \int_{t_0}^t \hat{H}_R(t') dt' + \mathcal{O}(\Delta^2).
\end{eqnarray}
This approach is valid when the driving conditions are far from adiabaticity, i.e. $\Delta^2/  v \ll 1$.
The probability $P(t) = \vert \Bra{\uparrow}\hat{U}(t, t_0)\Ket{\downarrow}\vert ^2$ to find the system in the excited state $\Ket{\uparrow}$ at time $t$ given that it started in the ground state $\Ket{\downarrow}$ at $vt_0 \ll -\Delta$ is given by
\begin{eqnarray}
  P(t) = \frac{\Delta^2}{4}\left\vert\int_{t_0}^{t} e^{i v t'^2/2}\,dt'\right\vert^2
\end{eqnarray}
as given in Eq. (2) of the main text.
The oscillatory dependence of this this integral on the final time $t$ can be verified with the help of the Cornu spiral.

\begin{figure}
\center{
\includegraphics[width=2.2in]{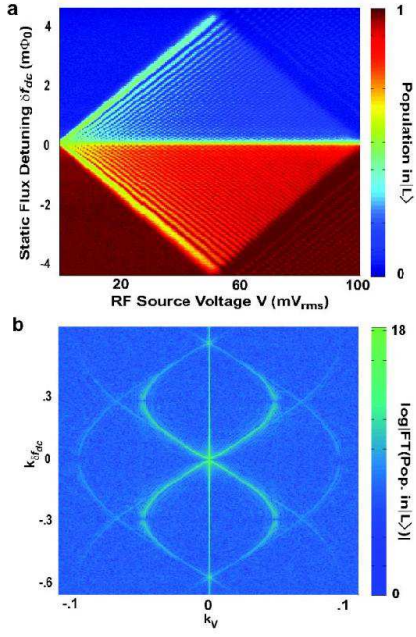}
\caption[t]{\textbf{Diamond 1 and its 2D Fourier transform. a,}
Diamond 1 in Fig.1 in the main text. Driving frequency $\nu = 160$ MHz. \textbf{b,} 2D Fourier
transform of diamond 1. A single sinusoid is visible, along with its lower harmonics~\cite{Rudner08a}.}}
 \label{OSMfig2}
\end{figure}
\begin{figure}
\vspace{0.2in}
\center{
\includegraphics[width=2.2in]{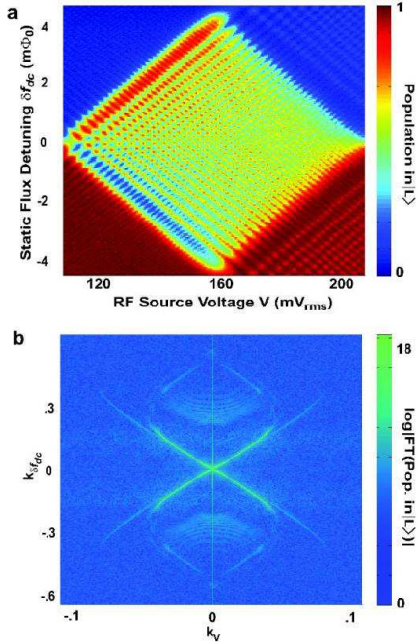}
\caption[t]{\textbf{Diamond 2 and its 2D Fourier transform. a,}
Diamond 2 in Fig.1 in the main text. Driving frequency $\nu = 160$ MHz. \textbf{b,} 2D Fourier
transform of diamond 2. Two sinusoids are visible with slightly different period due to the multiple energy bands and avoided crossings
that lead to diamond 2~\cite{Rudner08a}.}}
 \label{OSMfig3}
\end{figure}
\begin{figure}
\center{
\includegraphics[width=2.2in]{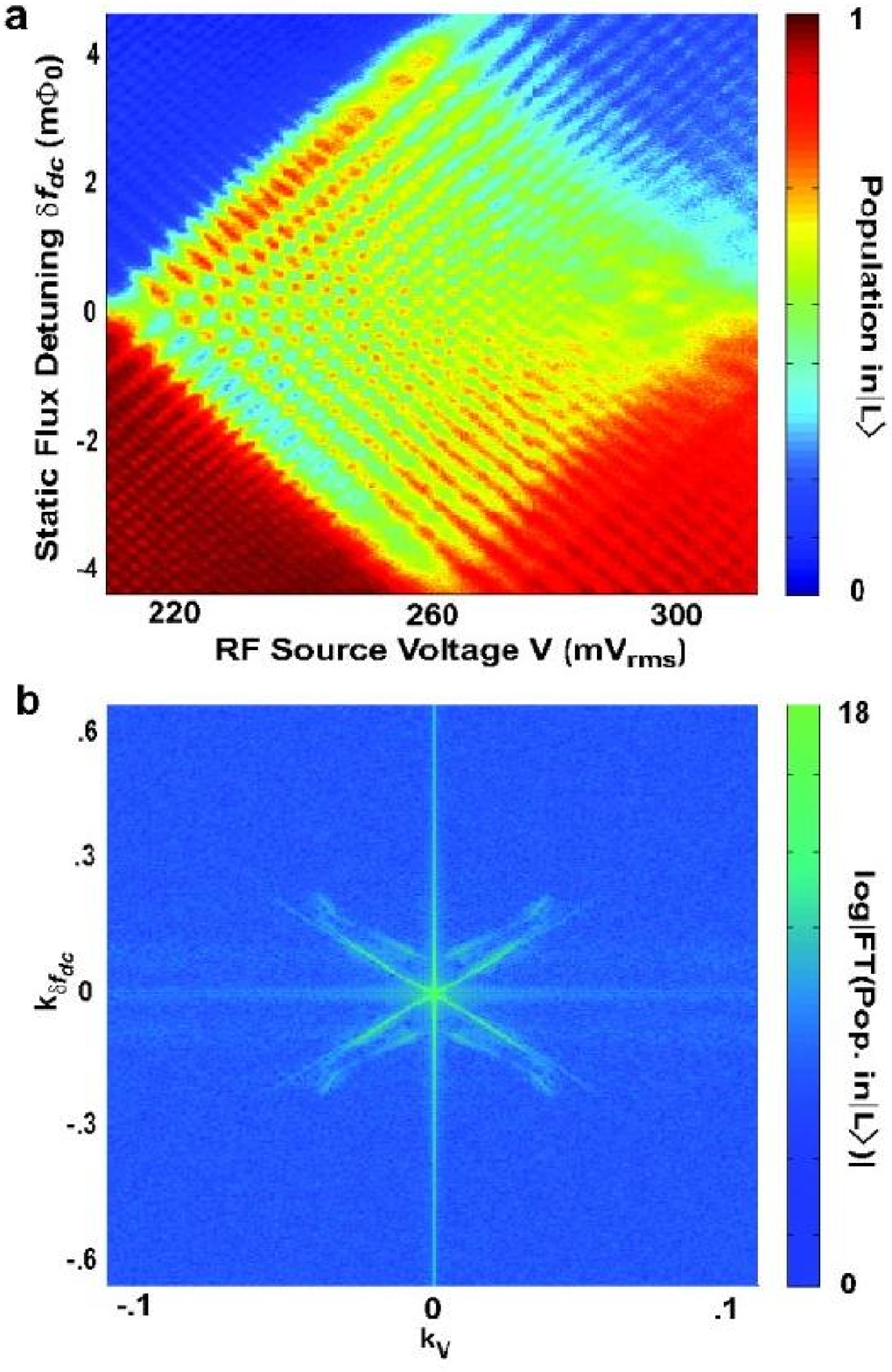}
\caption[t]{\textbf{Diamond 3 and its 2D Fourier transform. a,}
Diamond 3 in Fig.1 in the main text. Driving frequency $\nu = 160$ MHz. \textbf{b,} 2D Fourier
transform of diamond 3. Because the Fourier transform samples progressively smaller sectors of the diamonds as diamond number increases, the extent of the sinusoids in the $k_{\delta f_{\textrm{dc}}}$ direction is limited~\cite{Rudner08a}.}}
 \label{OSMfig4}
\end{figure}
\begin{figure}
\center{
\includegraphics[width=2.2in]{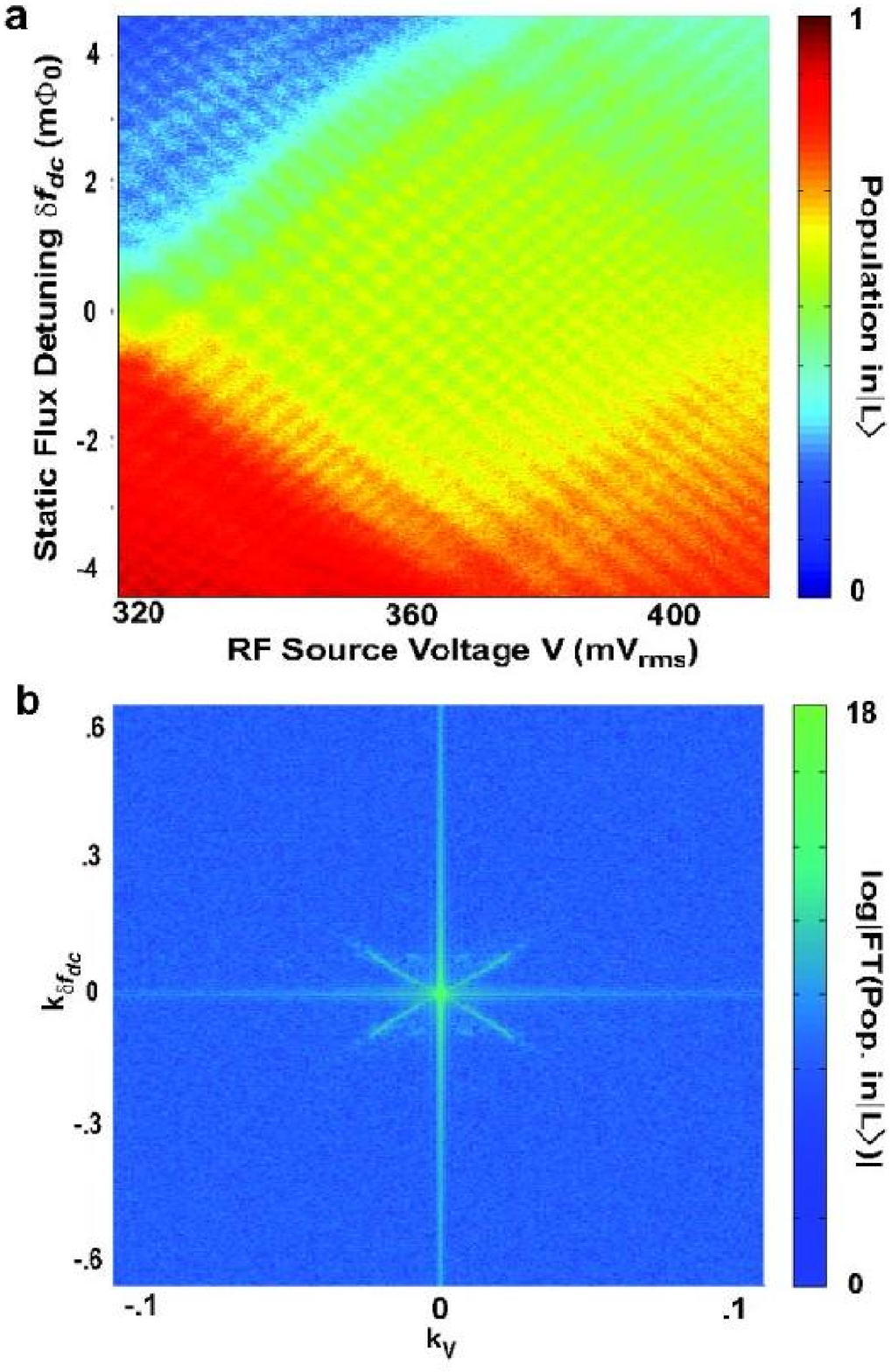}
\caption[t]{\textbf{Diamond 4 and its 2D Fourier transform. a,}
Diamond 4 in Fig.1 in the main text. Driving frequency $\nu = 160$ MHz. \textbf{b,} 2D Fourier
transform of diamond 4. Because the Fourier transform samples progressively smaller sectors of the diamonds as diamond number increases, the extent of the sinusoids in the $k_{\delta f_{\textrm{dc}}}$ direction is limited~\cite{Rudner08a}.}}
 \label{OSMfig5}
\end{figure}
%
\begin{figure}
\center{
\includegraphics[width=3.5in]{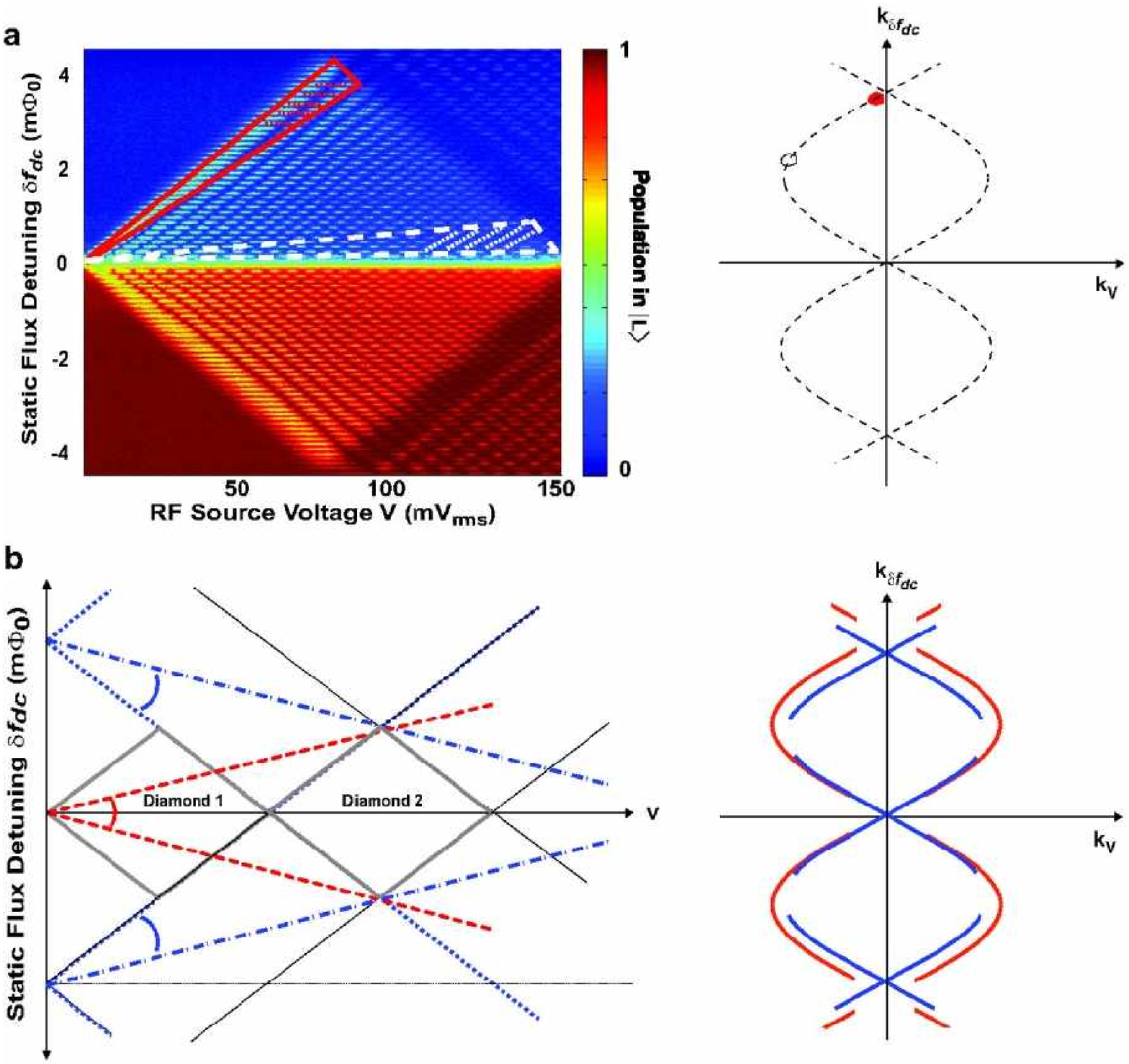}
\caption[t]{\textbf{Graphical interpretation of 2D Fourier transform technique.
a,}
Diamond 1 at $\nu=400$ MHz is pictured alongside a schematic of its Fourier transform.
On a local scale within the wedge-shaped region of $(V, \delta f_{\rm{dc}})$ space outlined by a solid red line, the image everywhere looks like a simple series of evenly spaced horizontal bands.
The Fourier transform over this region maps to a region of $(k_V, k_{\delta f_{\rm{dc}}})$ space localized near the $k_{\delta f_{\rm{dc}}}$ axis as indicated by the red dot in (B).
Within the region outlined by the dashed white line, the local periodicity is along the angled interference fringes; the Fourier transforms maps this region to the region of $(k_V, k_{\delta f_{\rm{dc}}})$ space localized near the extrema of the sinusoid in the $k_V$ direction as indicated by the open white circle.
\textbf{b,} The
Fourier integral samples a smaller sector of the real space image associated with the higher excited states in higher diamonds.  Due to the mapping between sectors of real space and localized regions of $k$-space, the sinusoids associated with higher diamonds are not fully developed. Only the portion of the sinusoid corresponding to the mapped region in real space is realized in the Fourier image.}}
 \label{OSMfig6}
\end{figure}

\begin{figure}
\center{
\includegraphics[width=2.7in]{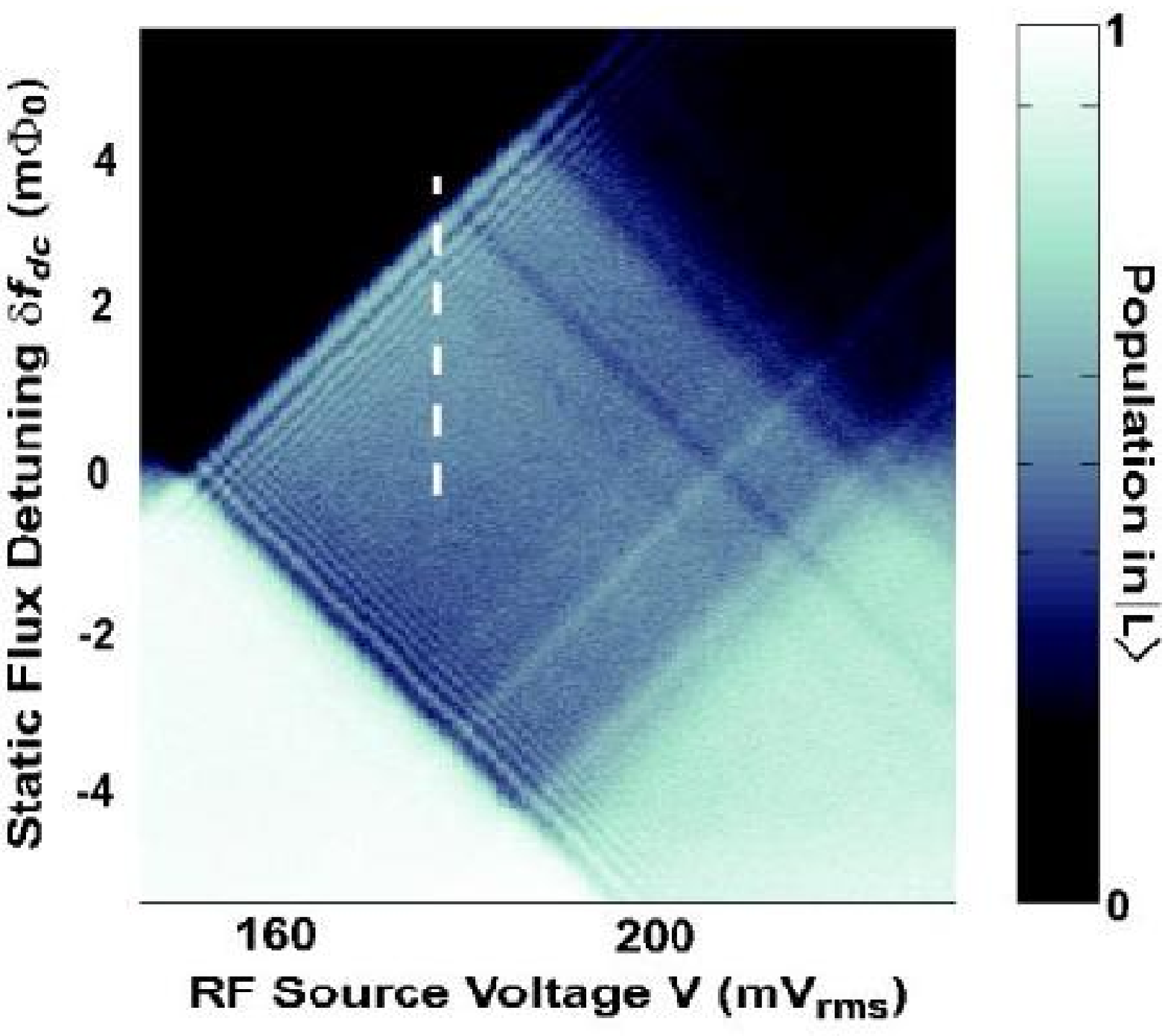}
\caption{\textbf{Diamond 3 at $\nu=45$ MHz.}  Dashed line indicates the amplitude ($181 \textrm{mV}_{\textrm{rms}}$) at
which the pulse-width scan in Fig.3 was taken.}}
 \label{OSMfig7}
\end{figure}

\begin{figure}
\center{
\includegraphics[width=2.9in]{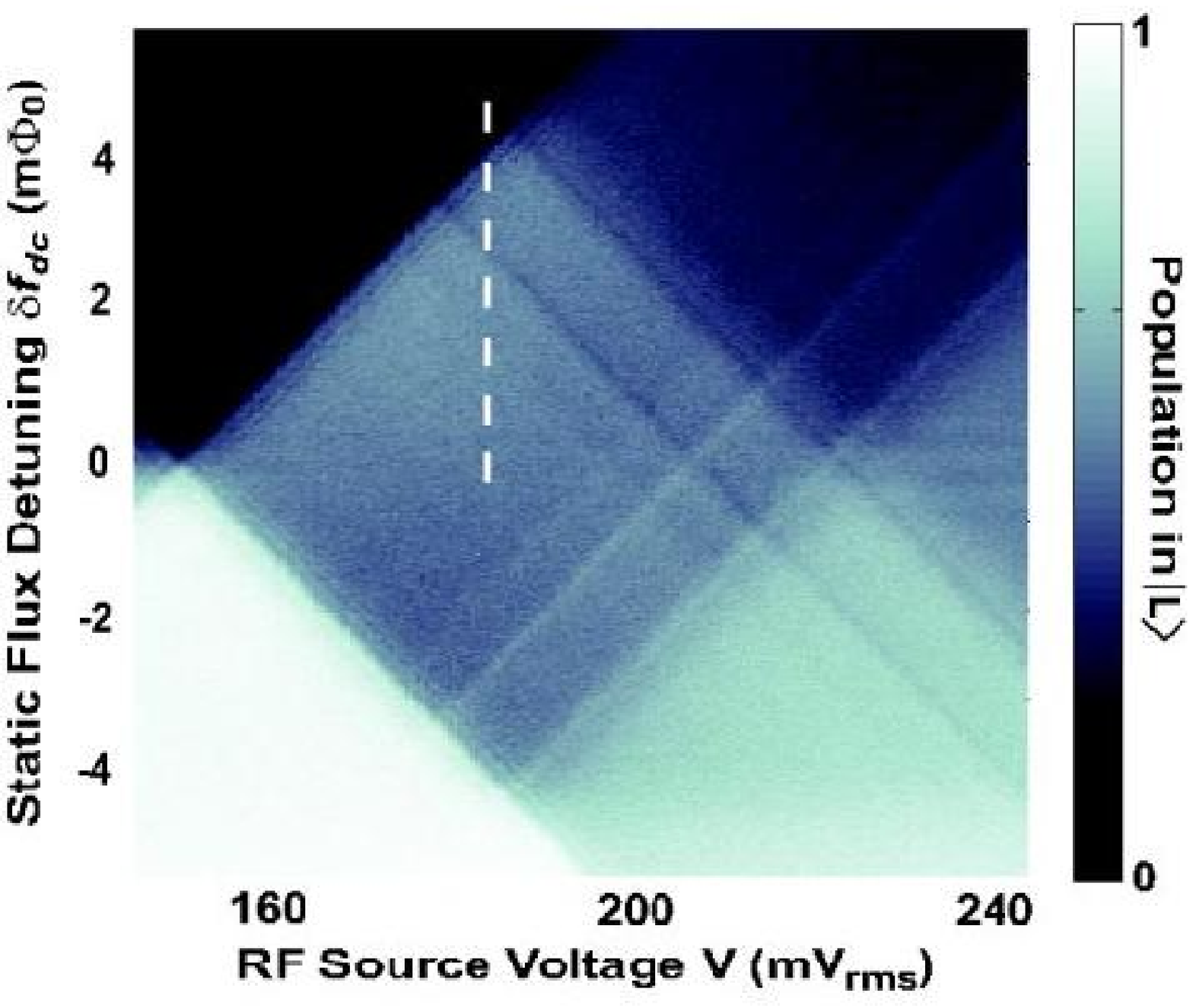}
\caption[t]{\textbf{Diamond 3 at $\nu=25$ MHz.}  Dashed line indicates the amplitude ($179 \textrm{mV}_{\textrm{rms}}$) at which
the pulse width scan in Fig.4 was taken.}}
 \label{OSMfig8}
\end{figure}

\begin{figure}
\center{
\includegraphics[width=2.9in]{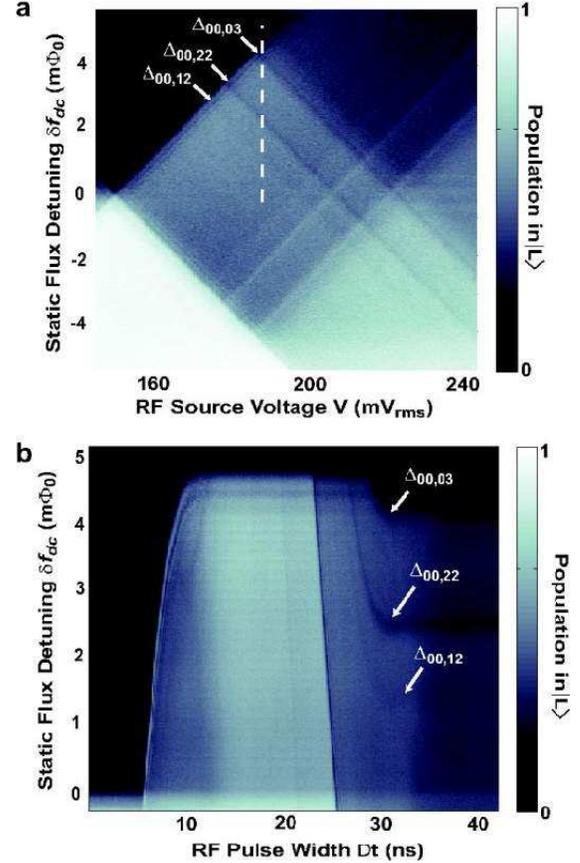}
\caption[t]{\textbf{Pulse width scan at $\nu=25$ MHz in diamond 3.
a,} Diamond 3 at $\nu=25$ MHz.  Dashed line indicates the amplitude ($183 \textrm{mV}_{\textrm{rms}}$) at which the pulse width scan in Fig.A13b was taken.
\textbf{b,}
Pulse width scan in diamond 3 at $\nu=25$ MHz. A loss
of population occurs where the crossings $\Delta_{00,03}$,
$\Delta_{22,00}$ and $\Delta_{12,00}$ are reached. This is in
contrast to Fig.4A, where crossing $\Delta_{03,00}$ was reached,
but never $\Delta_{00,03}$.
Also noted are the avoided crossings with which the diagonal lines in the diamond (see a) are attributed.}
}
 \label{OSMfig9}
\end{figure}

\clearpage
\bibliographystyle{Science}

\begin{thebibliography}{10}



\bibitem{Schawlow82a}
A.L. Schawlow,
\emph{Rev. Mod. Phys.} \textbf{54}, 697-707 (1982).

\bibitem{Thompson85}
R.C. Thompson,
\emph{Rep. Prog. Phys.} \textbf{48}, 531-578 (1985).



\bibitem{Friedman00a}
J.~R. Friedman,
V. Patel, W. Chen, S. K. Tolpygo, J. E. Lukens,
\emph{Nature} \textbf{406}, 43-46 (2000).

\bibitem{Wal00a}
C.~H. van der Wal, A. C. J. ter Haar, F. K. Wilhelm, R. N. Schouten,
C. J. P. M. Harmans, T. P. Orlando, S. Lloyd, J. E. Mooij,
\emph{Science} \textbf{290}, 773-777 (2000).

\bibitem{Berkley03a}
A.~J. Berkley, H. Xu, R. C. Ramos, M. A. Gubrud, F. W. Strauch,
P. R. Johnson, J. R. Anderson, A. J. Dragt, C. J. Lobb, F. C. Wellstood,
\emph{Science} \textbf{300}, 1548-1550 (2003).

\bibitem{vanderWiel03a}
W. G. van der Wiel,
\emph{et~al.},
\emph{Rev. Mod. Phys.} \textbf{75}, 1-22 (2003).

\bibitem{Clarke88}
J. Clarke, A. N. Cleland, M. H. Devoret, D. Esteve, J. H. Martinis,
\emph{Science} \textbf{239}, 992-997 (1988).

\bibitem{Hanson07a}
R. Hanson, L.P. Kouwenhoven, J.R. Petta, S. Tarucha, and L.M.K. Vandersypen,
\emph{Rev. Mod. Phys.} \textbf{79}, 1217-1265 (2007).


\bibitem{Nakamura99a}
Y.~Nakamura, Y.~A. Pashkin, J.~S. Tsai,
\emph{Nature} \textbf{398},
786 (1999).

\bibitem{Vion02a}
D.~Vion, A. Aassime, A. Cottet, P. Joyez, H. Pothier, C. Urbina, D. Esteve, and M. H. Devoret,
\emph{Science} \textbf{296}, 886-889 (2002).

\bibitem{Yu02a}
Y.~Yu, S. Han, X. Chu, S.-I. Chu, Z. Wang,
\emph{Science} \textbf{296}, 889-892 (2002).

\bibitem{Martinis02a}
J.~M. Martinis, S.~Nam, J.~Aumentado, C.~Urbina,
\emph{Phys. Rev. Lett.} \textbf{89}, 117901 (2002).

\bibitem{Chiorescu03a}
I.~Chiorescu, Y.~Nakamura, C.~J. P.~M. Harmans, J.~E. Mooij,
\emph{Science} \textbf{299}, 1869-1871 (2003).


\bibitem{Collin}
R.E. Collin
\emph{Foundations for microwave engineering} (Wiley-IEEE Press New York, 2001)




\bibitem{Nakamura01}
Y.~Nakamura, Y.~A. Pashkin, J.~S. Tsai,
\emph{Phys. Rev. Lett.} \textbf{87}, 246601 (2001).



\bibitem{Claudon04a}
J.~Claudon, F.~Balestro, F.~W.~J. Hekking, O.~Buisson,
\emph{Phys. Rev. Lett.}
\textbf{93}, 187003 (2004).

\bibitem{Plourde05a}
B. L. T. Plourde, T.L. Robertson, P.A. Reichardt, T. Hime, S. Linzen, C.-E. Wu, and John Clarke,
\emph{Phys. Rev. B}
\textbf{72}, 060506(R) (2005).

\bibitem{Saito06a}
S.~Saito,
T. Meno, M. Ueda, H. Tanaka, K. Semba, and H. Takayanagi,
\emph{Phys. Rev. Lett.} \textbf{96}, 107001 (2006).


\bibitem{Izmalkov04a}
A. Izmalkov, M. Grajcar, E. Il'ichev, N. Oukhanski, Th. Wagner, H.-G. Meyer, W. Krech, M. H. S. Amin, A. Maassen van den Brink and A. M. Zagoskin,
\emph{Europhys. Lett.} 65, 844-849 (2004).

\bibitem{Oliver05a}
W. D. Oliver, Y. Yu, J. C. Lee, K. K. Berggren, L. S. Levitov, T. P. Orlando,
\emph{Science} 310, 1653-1657 (2005).

\bibitem{Sillanpaa06a}
M. Sillanpaa, T. Lehtinen, A. Paila, Yu. Makhlin, P. Hakonen,
\emph{Phys. Rev. Lett.} 96, 187002 (2006).

\bibitem{Berns06a}
D.M. Berns, W.D. Oliver, S.O. Valenzuela, A.V. Shytov, K.K. Berggren, L.S. Levitov, T.P. Orlando,
\emph{Phys. Rev. Lett.} 97, 150502 (2006).

\bibitem{Wilson07a}
C.M. Wilson, T. Duty, F. Persson, M. Sandberg, G. Johansson, and P. Delsing,
\emph{Phys. Rev. Lett.} 98, 257003 (2007).


\bibitem{Valenzuela06a}
S. O. Valenzuela, W. D. Oliver, D. M. Berns, K. K. Berggren, L. S. Levitov, T. P. Orlando,
\emph{Science} 314, 1589-1592 (2006).

\bibitem{Niskanen07b}
A.O. Niskanen, Y. Nakamura, and J.P. Pekola,
\emph{Phys. Rev. B} {\bf 76}, 174523 (2007).

\bibitem{You08a}
J.Q. You, Yu-xi Liu, and Franco Nori,
\emph{Phys. Rev. Lett.} 100, 047001 (2008).



\bibitem{Chiorescu04a}
I. Chiorescu,
P. Bertet, K. Semba, Y. Nakamura, C.J.P.M. Harmans, and J.E. Mooij,
\emph{Nature} {\bf 431}, 159 (2004).

\bibitem{Wallraff04a}
A. Wallraff,
D. I. Schuster, A. Blais, L. Frunzio, R.-S. Huang,
J. Majer, S. Kumar, S. M. Girvin and R. J. Schoelkopf,
\emph{Nature} {\bf 431}, 162 (2004).

\bibitem{Johansson06a}
J.~Johansson,
S. Saito, T. Meno, H. Nakano, M. Ueda, H. Tanaka, K. Semba, and H. Takayanagi,
\emph{Phys. Rev. Lett.} \textbf{96}, 127006 (2006).


\bibitem{Makhlin01a}
Y.~Makhlin, G.~Sch\"on, A.~Shnirman,
\emph{Rev. Mod. Phys.} \textbf{73}, 357 (2001).

\bibitem{Mooij05}
J. E. Mooij, The Road to Quantum Computing, Science {\bf 307}, 1210 (2005)


\bibitem{Hime06a} T. Hime, 
    P.A. Reichart, B.L.T. Plourde, T.L. Robertson, C.-E. Wu, A.V. Ustinov, and J. Clarke,
    \emph{Science} {\bf 314} 1427 (2006)

\bibitem{Ploeg07a}
    S.H.W. van der Ploeg, 
    A. Izmalkov, A. Maassen van den Brink, U. H\"{u}bner, M.
    Grajcar, E. Il'ichev, H.-G. Meyer, and A.M. Zagoskin,
    \emph{Phys. Rev. Lett.} {\bf 98}, 057004 (2007).

\bibitem{Niskanen07a}
    A.O. Niskanen, K. Harrabi, F. Yoshihara, Y. Nakamura, S. Lloyd, and J.S. Tsai,
    \emph{Science} {\bf 316}, 723 (2007).


\bibitem{Sillanpaa07a}
M.A. Sillanpaa, J.I. Park, and R.W. Simmonds,
\emph{Nature} {\bf 449}, 438 (2007).

\bibitem{Majer07a}
J. Majer, J.M. Chow, J.M. Gambetta, J. Koch, B.R. Johnson, J.A. Schreier, L. Frunzio, D.I. Schuster,
A.A. Houck, A. Wallraff, A. Blais, M.H. Devoret, S.M. Girvin, and R.J. Schoelkopf
\emph{Nature} {\bf 449}, 443 (2007).


\bibitem{Pashkin03a}
Y.~A. Pashkin, T. Yamamoto, O. Astafiev, Y. Nakamura, D. V. Averin, and J. S. Tsai,
\emph{Nature} \textbf{421}, 823 (2003).

\bibitem{Yamamoto03a}
T.~Yamamoto, Yu. A. Pashkin, O. Astafiev, Y. Nakamura, J. S. Tsai,
\emph{Nature} \textbf{425}, 941 (2003).

\bibitem{McDermott05a}
R.~McDermott, R. W. Simmonds, M. Steffen, K. B. Cooper, K. Cicak,
K. D. Osborn, S. Oh, D. P. Pappas, J. M. Martinis,
\emph{Science} \textbf{307}, 1299 (2005).

\bibitem{Plantenberg07a}
J.H. Plantenberg, P.C. de Groot, C.J.P.M. Harmans, and J.E. Mooij,
\emph{Nature} \textbf{447}, 836 (2007).


\bibitem{Steffen06a}
M. Steffen, A. Ansmann, R.C. Bialczak, N. Katz, E. Lucero, R.~McDermott, M. Neeley, E.M. Weig, A.N. Cleland, and J. M. Martinis,
\emph{Science} \textbf{313}, 1423 (2006).


\bibitem{Siddiqi04a}
I. Siddiqi, R. Vijay, F. Pierre, C.M. Wilson, M. Metcalfe,
C. Rigetti, L. Frunzio, and M.H. Devoret,
\emph{Phys. Rev. Lett.} {\bf 93}, 207002 (2004)

\bibitem{Katz06a}
N. Katz, M. Ansmann, R.C. Bialczak, E. Lucero, R. McDermott, M. Neeley, M. Steffen,
E.M. Weig, A.N. Cleland, J.M. Martinis, A.N. Korotkov
\emph{et~al.},
\emph{Science} {\bf 312}, 1498 (2006)

\bibitem{Lupascu07a}
A. Lupascu, S. Saito, T. Picot, P.C. de Groot, C.J.P.M. Harmans, J.E. Mooij,
\emph{Nature Physics} {\bf 3}, 119 (2007)


\bibitem{Mooij99a}
J.~E.~Mooij, T.~P.~Orlando, L.~S.~Levitov, L.~Tian, C.~H.~{van~der~Wal},
 S.~Lloyd,
\emph{Science} \textbf{285},
1036-1039 (1999).


\bibitem{Orlando99a}
T.~P.~Orlando, J.~E.~Mooij, L.~Tian, C.~H.~{van~der~Wal},
L.~S.~Levitov, S.~Lloyd, and J.~J.~Mazo,
\emph{Phys. Rev.~B} \textbf{60},
15398-15413 (1999).






\bibitem{H_Nakamura01}
H. Nakamura, \emph{Nonadiabatic Transition} (London, England:
World Scientific, 2001).


\bibitem{Mark07a}
M. Mark, T. Kraemer, P. Waldburger, J. Herbig, C.Chin, H.-C. N\"{a}gerl, and R. Grimm
\emph{Phys. Rev. Lett.} {\bf 99}, 113201 (2007).


\bibitem{Astafiev07a}
O. Astafiev, K. Inomata, A.O. Niskanen, T. Yamamoto, Yu. A. Pashkin, Y. Nakamura, and J.S. Tsai,
\emph{Nature} {\bf 449}, 588 (2007).


\bibitem{Rudner08a}
M.S. Rudner
\emph{et~al.},
%
arXiv:0805.1555.


\bibitem{Mark07b}
M. Mark, F. Ferlaino, S. Knoop, J.G. Danzl, T. Kraemer, C.Chin, H.-C. N\"{a}gerl, and R. Grimm,
\emph{Phys. Rev. A} {\bf 76}, 042514 (2007).

\bibitem{Lang08a}
F. Lang, P.V.D. Straten, B. Brandst\"{a}tter, G. Thalhammer, K. Winkler, P.S. Julienne, R. Grimm,
and J. Hecker Denschlag,
\emph{Nature Physics} {\bf 4}, 223-226 (2008).

\bibitem{Ashab07a}
S. Ashab, J.R. Johansson, A.M. Zagoskin, and Franco Nori,
\emph{Phys. Rev. A} {\bf 75}, 063414 (2007).

\end{thebibliography}

\end{document}